\documentclass[journal]{IEEEtran}                                               \usepackage{cite}
\usepackage{amsmath,amssymb,amsfonts}
\usepackage{algorithmic}
\usepackage{graphicx}
\usepackage{textcomp}
\usepackage{lineno}

\def\BibTeX{{\rm B\kern-.05em{\sc i\kern-.025em b}\kern-.08em
    T\kern-.1667em\lower.7ex\hbox{E}\kern-.125emX}}

\begin{document}
\title{Spectral Ultrasound Imaging of Speed-of-Sound and Attenuation Using an Acoustic Mirror}
\author{Bhaskara Rao Chintada, Richard Rau, and Orcun Goksel, 
\thanks{B.R. Chintada, R. Rau, and O. Goksel are with the Computer-assisted Applications in Medicine (CAiM) group, ETH Zurich, Switzerland.}
\thanks{O Goksel is also with the Department of Information Technology, Uppsala University, Sweden. (email: ogoksel@ethz.ch)}}
\maketitle
\begin{abstract}
Speed-of-sound and attenuation of ultrasound waves vary in the tissues.
There exist methods in the literature that allow for spatially reconstructing the distribution of group speed-of-sound (SoS) and frequency-dependent ultrasound attenuation (UA) using reflections from an acoustic mirror positioned at a known distance from the transducer.
These methods utilize a conventional ultrasound transducer operating in pulse-echo mode and a calibration protocol with measurements in water.
In this study, we introduce a novel method for reconstructing local SoS and UA maps as a function of acoustic frequency through Fourier-domain analysis and by fitting linear and power-law dependency models in closed form.
Frequency-dependent SoS and UA together characterize the tissue comprehensively in spectral domain within the utilized transducer bandwidth. 
In simulations, our proposed methods are shown to yield low reconstruction error: 0.01\,dB/cm$\cdot$MHz$^\mathbf{y}$ for attenuation coefficient and 0.05 for the frequency exponent. 
For tissue-mimicking phantoms and \textit{ex-vivo} bovine muscle samples, a high reconstruction contrast was achieved.
Attenuation exponents in a gelatin-cellulose mixture and an \textit{ex-vivo} bovine muscle sample were found to be, respectively, 1.3 and 0.6 on average.
Linear dispersion of SoS in a gelatin-cellulose mixture and an \textit{ex-vivo} bovine muscle sample were found to be, respectively, 1.3 and 4.0\,m/s$\cdot$MHz on average.
These findings were reproducible when the inclusion and substrate materials were exchanged.
Bulk loss modulus in the bovine muscle sample was computed to be approximately 4 times the bulk loss modulus in the gelatin-cellulose mixture.
Such frequency-dependent characteristics of SoS and UA, and bulk loss modulus may therefore differentiate tissues as potential diagnostic biomarkers.
\end{abstract}


\section{Introduction}
\label{sec:introduction}
Medical imaging methods aim to characterize and spatially map different soft tissue properties in order to provide diagnostic information regarding pathological structures and processes.
Ultrasound imaging is a cost-effective, real-time and non-ionizing medical imaging modality. 
Conventional B-mode ultrasound imaging aims to map the amplitude of ultrasound waves scattered and reflected from tissue structures.
Complementary to this, several methods exist in the literature to quantify various tissue characteristics.
For instance, shear-wave elastography imaging (SWEI) aims to infer local soft tissue shear moduli, often derived from the group speed of propagating shear-waves that are induced using acoustic radiation force pushes~\cite{catheline2004measurement,bercoff2004supersonic}.
Since soft tissues are inherently viscoelastic~\cite{fung2013biomechanics}, several ultrasound-based imaging techniques were proposed to fully characterize the soft tissues in spectral domain by imaging the shear-wave speed~\cite{chen2004quantifying,deffieux2008shear,chen2009shearwave} and shear-wave attenuation as a function of frequency~\cite{nenadic2016attenuation,budelli2017diffraction,bernard2016frequency}, which then help to determine the shear and storage moduli of the medium.
Methods also exist in the shear-wave literature that aim to characterize the nonlinear response of soft tissues given elastic~\cite{gennisson2007acoustoelasticity,latorre2012quantitative,otesteanu2017quantification,bernal2016vivo,chintada2019acoustoelasticity} or viscoelastic~\cite{otesteanu2019spectral,chintada2020nonlinear,goswami2020local} constitutive models and assumptions.

Inspired by such comprehensive mechanical characterization of soft tissue using group speed, phase velocity, and attenuation of \emph{shear waves} as a function of frequency, we herein propose to extend similar characterization to ultrasound waves themselves.
Ultrasound is a longitudinal (compressional) mechanical wave, and through its propagation in tissues, its characteristics also change based on tissue mechanical properties.
Accordingly, ultrasonic group speed, phase velocity, and attenuation as a function of frequency can be used for further characterization of soft tissues.
Indeed, ultrasound group speed (commonly called speed-of-sound, SoS) is related to medium bulk modulus while phase velocity and attenuation are related to bulk storage and loss moduli. 
Imaging of (group) SoS has been studied in the literature using tomographic approaches with custom made ultrasound transducer and data acquisition solutions in~\cite{duric2007detection,greenleaf1981clinical,gemmeke20073d}, which involve bulky and costly setups and submersion of the imaged anatomy in water. 
To utilize the practical advantages of commercial transducer arrays, SoS imaging using conventional transducers in pulse-echo mode has been proposed recently, following two approaches: A group of methods measure apparent displacements of backscattered signals when insonified from different angles~\cite{jaeger2015computed,sanabria2018spatial,rau2021diverging}.
Alternative methods use an additional passive acoustic reflector on the opposite side of the imaged anatomy to measure several time-of-flight values from which to tomographically reconstruct local SoS distribution~\cite{krueger1998limited,sanabria2018speed}.

Similarly to SoS imaging, imaging ultrasound attenuation (UA) was also proposed using custom-made water-submersion setups~\cite{duric2007detection}.
Using conventional ultrasound transducers, some methods utilize the fact that the ultrasound center frequency decreases as a function of propagation depth and medium UA, since attenuation affects higher frequencies more prominently.
Based on this, a spectral difference method~\cite{parker1983measurement} and spectral log difference methods~\cite{coila2017regularized,vajihi2018low} were proposed using typical B-mode images together with reference phantom measurements.
An adjacent frequency normalization was proposed in~\cite{gong2019system} to cancel out systemic effects such as focusing and time-gain-compensation without the need of reference phantom measurements.
Alternatively, reference measurements from a passive acoustic mirror was used in~\cite{huang2005ultrasonic,chang2007reconstruction} to estimate UA values, however this requires apriori segmentation masks to be given and hence do not allow the UA imaging of arbitrary unknown domains.
Recently, imaging of local UA has been proposed in~\cite{Rau_attenuation_19} using a limited-angle computed tomography (LA-CT) method.

Formally, UA $\alpha(\omega)$ is a function of ultrasound frequency $\omega$ and it typically follows a power-law relation of the form $\alpha(\omega)=\alpha_0 \omega^y$, where $\alpha_0$ is called the attenuation coefficient and $y$ the power exponent.
The above-mentioned methods in the literature either estimate a single UA map, e.g., at the ultrasound transmit center frequency or estimate the attenuation coefficient $\alpha_0$ map assuming $y=1$.
The latter assumption of $y\approx1$ is not necessarily valid  in general as it was shown that soft tissues exhibit varying UA power exponents $y$~\cite{bamber1981acoustic,parker1983measurement}.
Furthermore, using a single (center) frequency for the analysis neglects the fact that the spectral composition of propagating waves change given the larger attenuation at higher frequencies.
Characterizing UA using a parametric model over a frequency range would enable a complete spectral characterization.
This would allow for treating the UA variation within the utilized bandwidth as information, which is treated as noise when the underlying model is neglected.
The parameters of such model can also yield additional imaging biomarkers characterizing the spectral tissue behaviour.
With this motivation, recently in~\cite{rau2021frequency} it was proposed to reconstruct multiple UA maps ($\alpha$) at different band-pass filtered frequency bands.
To estimate $\alpha_0$ and $y$ locally, these UA maps were then pixel-wise fitted with a power-law model by assuming the UA estimates to characterize the response at the central frequency of each band.
However, band-pass filtering the signal over a frequency band then averages the cumulative effect of the entire frequency band.
Also, the pixel-wise model fitting ignores the spatial continuity of the image, where the individual frequency band reconstructions cannot benefit from each other and the SNR loss due to dividing into separate pass-bands cannot be recovered.
A frequency-domain solution, similar to those in shear-wave applications and which takes into account the entire bandwidth in a single inverse problem formulation, could mitigate the above shortcomings and is proposed herein for UA reconstruction.

Similarly to UA, SoS also varies not only across tissues, e.g.\ in the human liver depending upon water, fat, and collagen concentrations~\cite{bamber1981acoustic}, but also across acoustic frequencies~\cite{mast1999simulation, sehgal1986measurement, mast2000empirical,treeby2011measurement}.
SoS is shown to vary with respect to frequency in soft tissues in the literature~\cite{duck2013physical}.
For instance, it was shown in~\cite{kremkau1981ultrasonic} that SoS varies with frequency in human brain tissues.
Thus, frequency-dependent characterization of SoS may be an additional biomarker for tissue characterization. 

In this study, we introduce a new method based on general wave theory in frequency domain to compute both SoS and UA across frequencies and compute the SoS and UA model parameters together using a inverse problem formulation.
We have conducted simulations and \textit{ex-vivo} phantom studies to discuss the reconstruction results. 

\section{Methods}

\begin{figure*}
    \centering
    \includegraphics[width=0.9\textwidth]{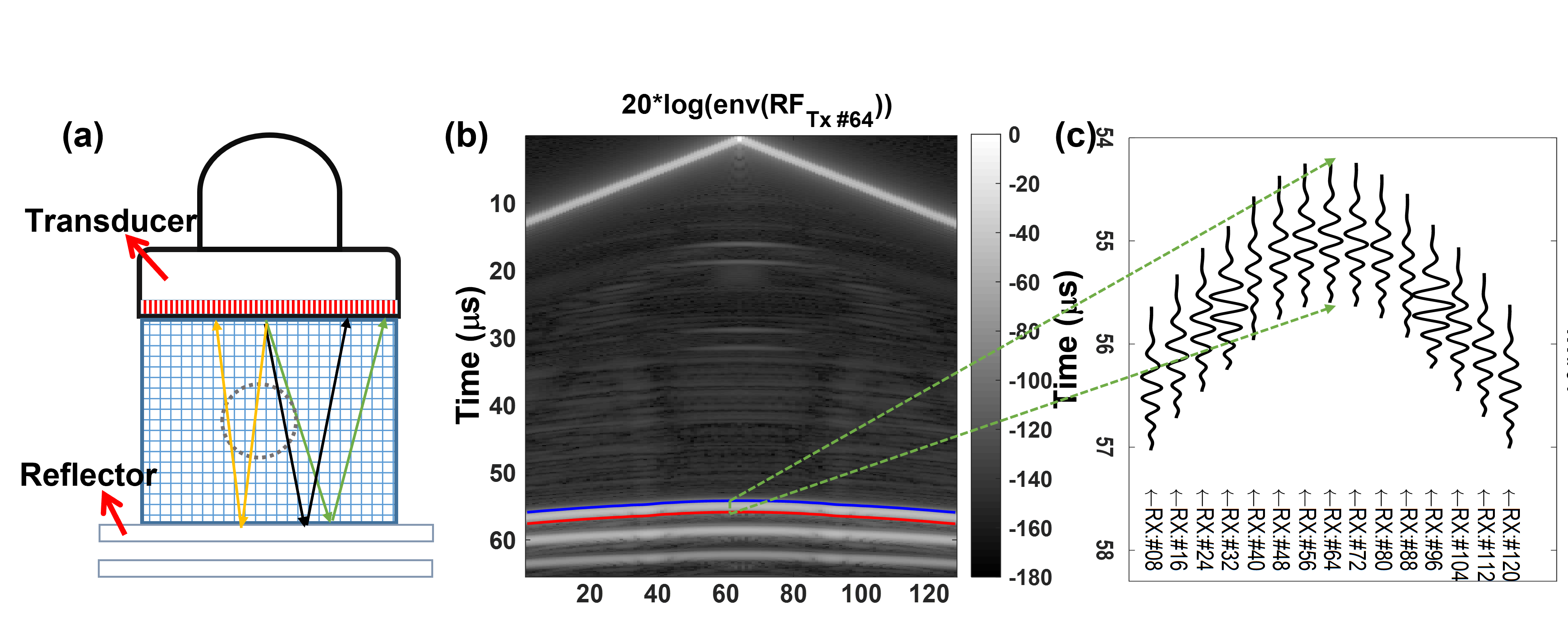}
    \caption{(a)~Schematic of the imaging setup where an acoustic mirror (reflector) is positioned at a pre-defined distance from the ultrasound transducer surface, with coloured paths representing the traversal of different acoustic wavefront paths from a single transmit element to different receivers through a discretized tissue representation, (b)~Pre-beamformed data received by all the receive channels of the transducer when the transducer element \#64 is used for transmission (Tx).
    (c)~Cropped reflector echo profiles corresponding to different Rx elements, from which the reflector is next delineated precisely.}.
    \label{fig:methods_1}
\end{figure*}

\begin{figure*}
    \centering
    \includegraphics[width=.9\textwidth]{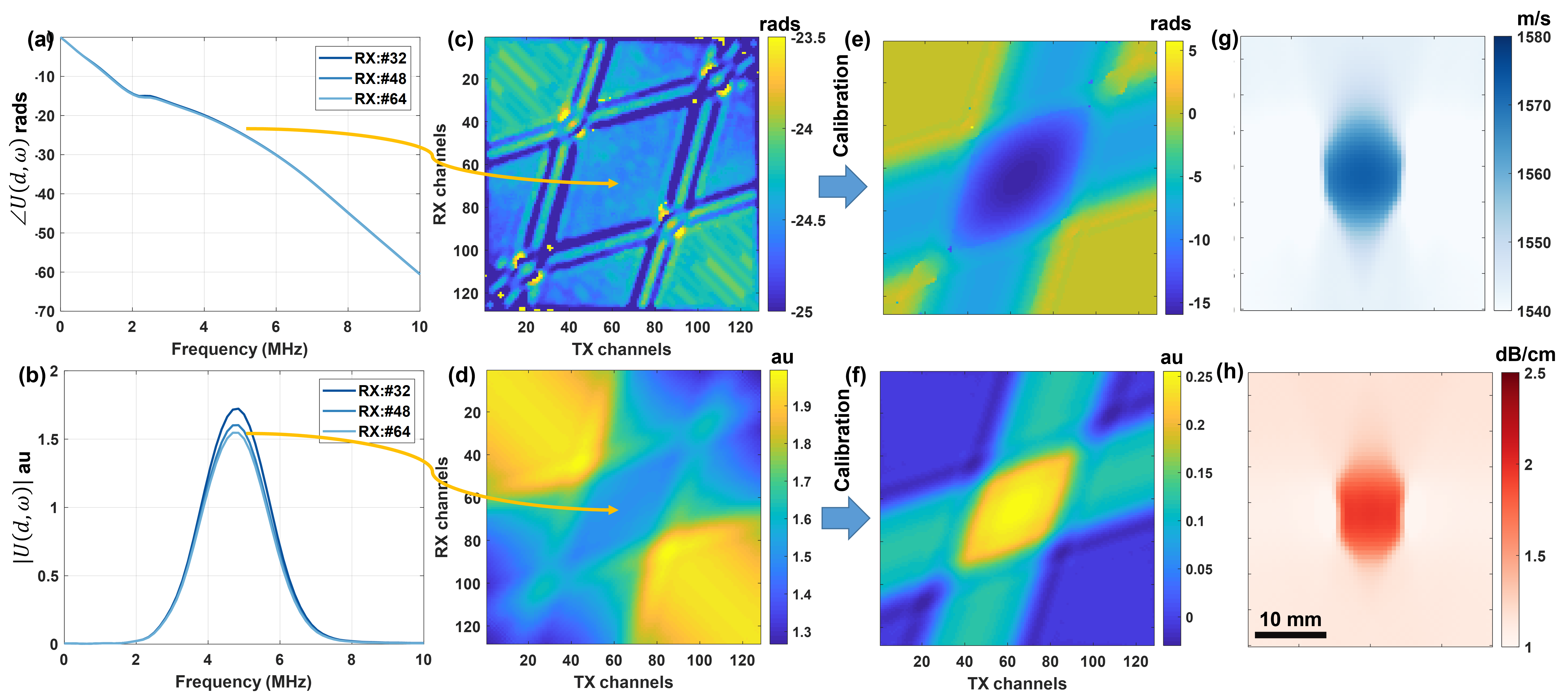}
    \caption{Procedure of estimating SoS and UA distribution maps at frequency $\omega$.
    1D temporal Fourier transform is performed on the echo profiles obtained at different Rx channels to obtain (a)~phase spectrum $\angle U(d,\omega)$ and (b)~amplitude spectrum $|U(d,\omega)|$ for three sample profiles shown in Fig.\,\ref{fig:methods_1}(c).
    Phase and amplitude spectrum values for all Tx-Rx combinations at a selected sample frequency of 5\,MHz are shown, respectively, in~(c) and~(d).
    These values are then calibrated as described in Sec.\,\ref{fwd_problem} leading to the calibrated phase~(e) and amplitude~(f) spectrum values, which are used for solving inverse problems to reconstruct local SoS~(g) and UA~(h) distributions at $\omega$.}
    \label{fig:SoS_UA_recon}
\end{figure*}

We acquire ultrasound data in multistatic mode, in which a single transducer element at a time is used for transmitting (Tx) a broadband ultrasound pulse into the medium and an acoustic mirror 
is placed at a predefined distance from the ultrasound transducer surface.
The reflected ultrasound echoes are recorded as received (Rx) with all the transducer elements.
This process is repeated until all the transducer elements are used for the transmission.
Multistatic data acquisition is schematically described in Fig.\,\ref{fig:methods_1}(a) for transmission with an element and receiving at different transducer elements with varying path lengths (marked with different colors). 
Full-matrix multistatic data is then processed using the algorithm in~\cite{chintada2021time} to delineate the reflector profile in the echoes reflected from the acoustic reflector positioned at a pre-defined depth.
A sample delineation for Tx\#64 of a 128 element transducer is marked on the pre-beamformed channel data in Fig.\,\ref{fig:methods_1}(b).
Other high amplitude echo profiles below the first arrival echoes seen in Fig.\,\ref{fig:methods_1}(b) correspond to reflections from different interfaces of the acoustic reflector, for more details please refer to Fig.\,1 in~\cite{chintada2021time}.
Sample cropped reflector echo profiles for Tx\#64 and Rx\#\{8,16,\ldots,120\} are plotted in Fig.\,\ref{fig:methods_1}(c).
For reconstruction of SoS and UA at a particular frequency we use phase spectrum and amplitude spectrum values, respectively, obtained by the Fourier transform (FT) of the above echo profiles around the delineated reflector time points. 
Phase and amplitude spectrums for some of the echo profiles shown in Fig.\,\ref{fig:methods_1} are plotted in Fig.\,\ref{fig:SoS_UA_recon}(a-b) respectively.
For reconstruction of SoS and UA at a particular frequency we use the amplitude and phase spectrum values at that frequency, for all combinations of Tx-Rx as shown in Fig.\,\ref{fig:SoS_UA_recon}(c-d).
These phase and amplitude spectrum readings for all Tx-Rx combinations are then corrected using pre-acquired calibration measurements in water, cf.\ Fig.\,\ref{fig:SoS_UA_recon}(e-f), which are used for solving inverse problems to reconstruct SoS and UA at that chosen frequency as shown in Fig.\,\ref{fig:SoS_UA_recon}(g-h).
The above method overview is elaborated and detailed in the following sections. 

\subsection{Ultrasound wave propagation in viscoelastic medium}
As a broadband ultrasound wave passes through a medium, its waveform changes according to the phase velocity and attenuation of the medium.
Ultrasound waves travelling through a viscoelastic medium can be defined in the frequency domain as:
\begin{subequations}
    \label{ultrasound_prop}
    \begin{align}
        \label{eq:ultrasound_prop_a}
         U(d,\omega) & = G(\omega) e^{-i\hat{k}(\omega)d}\\
        \label{eq:ultrasound_propb}
        \mathrm{with\quad} \hat{k}(\omega) & = \frac{\omega}{c(\omega)}-i\alpha(\omega)
    \end{align}
\end{subequations}
where $U(d,\omega)$ is the FT of the ultrasonic wave $u(d,t)$, $\omega$ is the angular frequency, $d$ is the propagation distance, $\hat{k}(\omega)$ is the complex wave number, $c(\omega)$ is the phase velocity, and $\alpha(\omega)$ is  attenuation at $\omega$, and
$G(\omega)$ accounts for the transmit and receive transducer responses at frequency $\omega$. 

As seen in Eq.\,\eqref{eq:ultrasound_prop_a}, ultrasonic wave amplitude decreases as a function of propagation distance $d$ due to absorption, relaxation, scattering, etc. 
Indeed using this wave equation, ultrasound phase velocity and attenuation can be derived using amplitude and phase spectrums by taking the natural logarithm of amplitude $|U(d,\omega)|$ and separating the phase terms to equate to phase angle $\angle U(d,\omega)$, which yield respectively:
\begin{subequations}
    \begin{align}
        \label{eq:phase_and_attenuation_eq}
        \angle U(d,\omega) & = {\frac{\omega}{c(\omega)}d+\angle G(\omega)},\\
        \ln (|U(d,\omega)|) & = \ln(G(\omega)) {-\alpha(\omega)d}.
    \end{align}
\end{subequations}
\subsection{Forward problem of ultrasound wave propagation}
\label{fwd_problem}
Phase  and amplitude spectra for the reflector echo corresponding to Tx element $t$ and Rx element $r$ relates, respectively, to \emph{slowness} $s_p(\omega)=1/c(\omega)$, i.e.\ the inverse of phase velocity, and attenuation $\alpha(\omega)$ along the traversed ray path $p$ as follows:
\begin{subequations}
    \label{fwd_relations}
    \begin{align}
    \label{slowness_fwd}
    \angle U_{t,r}(\omega) &= \omega \int_{p}
    \!s(\omega)\;\mathrm{d}l+\angle G_{t,r}(\omega,\theta_{t,r})\,,\\
    \label{attenuation_fwd}
    |U_{t,r}(\omega)| &= \exp\Big(-\int_{p} \!\alpha(\omega)\;\mathrm{d}l\Big)+|G_{t,r}(\omega,\theta_{t,r})|\,.
   \end{align}
\end{subequations}
where $p$ represents the acoustic ray path and $l$ is the traversed
distance.
$\angle G_{t,r}(\omega,\theta_{t,r})$ and $|G_{t,r}(\omega,\theta_{t,r})|$ are the confounding factors attributable to transducer Tx-Rx phase and amplitude transfer characteristics at different angles (e.g., the aperture opening and side-lobes) as well as the acoustic reflector reflection characteristics, i.e.\ the incidence angle dependent specular and diffuse reflection and transmission characteristics of the reflector surface.
These have to be compensated for in order to be able to measure the actual ultrasound phase velocity and attenuation effects.
For this purpose, we calibrate these measurements by normalizing with measurements in water ($U^{water}_{t,r}(\omega)$) at a known water temperature and using the same ultrasound pulses (bandwidth) as in the targeted tissue imaging setup.
From the given water temperature, we compute the ground-truth water SoS using the relationship in~\cite{del1972speed}, which was 1483.1\,m/s in our setup.
Water UA is known from~\cite{krautkramer2013ultrasonic} to be $\alpha(f)=2.17\times 10^{-15}f^2$.
 
We used the procedure described in~\cite{rau2021frequency} to compensate for the confounding effects in $|G_{t,r}(\omega,\theta_{t,r})|$.
The reflector does not change the phase of the reflected ultrasound signals, since its acoustic impedance is much higher than water and soft tissues.
Thus, for factoring out the confounding effects in $\angle G_{t,r}(\omega)$, we used simply the phase angles of reflector echo profiles from water calibration experiments.
We denote these calibrated amplitude and phase spectrum values as $|U_{t,r}'(\omega)|$ and $\angle U_{t,r}'(\omega)$, respectively, which are given by
\begin{subequations}
    \label{calibration}
    \begin{align}
    \label{amplitude_calibration}
    \angle U'_{t,r}(\omega) &=  \angle U_{t,r}(\omega) -\angle U^\mathrm{water}_{t,r}(\omega)\,\\
    \label{phase_calibration}
    |U'_{t,r}(\omega)| &= \frac{U_{t,r}(\omega)R_\mathrm{water}(\theta)}{U^\mathrm{water}_{t,r}(\omega)R(\theta)}.
   \end{align}
\end{subequations}
where $R_\mathrm{water}(\theta)$ and $R(\theta)$ are the ultrasound reflection coefficients of the reflector in water and in targeted imaging medium, respectively, for incident angle $\theta$.
It can be estimated using Snell's law with the SoS in water $(c_\mathrm{water})$ and an approximate SoS of the targeted medium, as in~\cite{rau2021frequency}.

For acoustic propagation, refractions are herein ignored and the travel of acoustic wave from a transmitting to receiving element is modeled using a straight ray approximation as depicted in Fig.\,\ref{fig:methods_1}a.
For a straight acoustic path, the first wavefront from a Tx element, reflecting from the reflector and arriving at an Rx element can be assumed to follow the shortest path, thus being reflected at the mid-point between these Tx and Rx element locations projected on the reflector, cf.\,Fig.\,\ref{fig:methods_1}(a).
To cast this as a tomographic reconstruction problem, the relation in Eqs.\,\ref{fwd_relations} above can be discretized on a Cartesian grid as 
\begin{subequations}
    \label{fwd_sums}
    \begin{align}
    \label{slowness_fwd_sum}
    \angle U'_{t,r}(\omega) &= \omega \sum_{n=1}^{N_i} {s}_k(\omega)\; l_{ik}\,,\\
    \label{attenuation_fwd_sum}
    |U'_{t,r}(\omega)| &= \exp\Big(-\sum_{n=1}^{N_i} {\alpha}_k(\omega)\;\mathrm{d}l\Big).
   \end{align}
\end{subequations}
where $s_k(\omega)$ and $\alpha_k(\omega)$ represents the slowness and attenuation, respectively, in pixel $k$ along ray $i$ spanning $N_i$ pixels, where $l_{ik}$ represents the partial acoustic path length of ray $i$ within this pixel $k$.

Collocating Eqs.\,\ref{fwd_sums} for each ray path along all Tx-Rx combinations and taking logarithm on both sides of Eq.\,\ref{attenuation_fwd_sum}, the forward-problem of ultrasound propagation can then be represented with the following systems of linear equations
\begin{subequations}
    \label{linear_forms}
    \begin{align}
    \label{slowness_linear_form}
    \omega\bf{L}s(\omega) &= \phi(\omega)\,,\\
    \label{attenuation_linear_form}
     \bf{L}\alpha(\omega) &= b(\omega)\,,
   \end{align}
\end{subequations}
where $\phi(\omega)$ and $b(\omega)$ are, respectively, the column vectors of phase $\angle U'_{t,r}(\omega)$ and amplitude $\ln|U'_{t,r}(r,\omega)|$ values along all ray paths at frequency $\omega$;
$s(\omega)$ and $\alpha(\omega)$ are the column vectors of, respectively, the slowness and attenuation values for all reconstruction grid pixels at frequency $\omega$; and the system matrix $\bf{L}$ encodes the discretized ray integrals for each ray onto the pixel grid.
Note that $\bf{L}$ is constant for given transducer geometry and reflector position.

\subsection{Reconstruction of phase velocity and attenuation}

Given calibrated phase spectrum $\phi(\omega)$ and amplitude spectrum $b(\omega)$ measurements, one can then reconstruct the local slowness and UA maps at frequency $\omega$ by solving the following inverse problems:
\begin{subequations}
    \label{eq:inv_forms}
    \begin{align}
    \label{eq:inv_sos}    
    \hat{s}(\omega) &=  \arg\min_{s(\omega)} {||\omega \bf{L}s(\omega)-\phi(\omega)||_1} + \lambda||Ds(\omega)||_1 \,\\
    \label{eq:inv_attenuation}
    \hat{\alpha}(\omega) &=  \arg\min_{\alpha (\omega)} {||\bf{L}\alpha(\omega)- b(\omega)||_1 }+ \lambda||D\alpha(\omega)||_1
    \end{align}
\end{subequations}
where $\lambda$ is the weight parameter of regularization required to help robustly solve these ill-posed problems.
Similarly to~\cite{sanabria2018spatial,sanabria2018speed,Rau_attenuation_19,rau2021frequency}, we herein use $l_\text{1}$-norm for both the data and regularization terms for robustness to outliers in, respectively, the measurements and the reconstructions~\cite{fu2006efficient}.
Due to a lack of full angular coverage of measurements, regularization matrix $D$ implements an LA-CT specific image filtering to suppress streaking artifacts orthogonal to missing projections via anisotropic weighting of directional gradients~\cite{sanabria2016hand}. 
Similarly to~\cite{sanabria2018speed}, we herein utilize a $k$=0.9 anisotropic weighting for a stronger regularization in the lateral direction, taking into account the uneven distribution of measurement paths. 
In this paper we empirically set $\lambda$=0.6 for all experiments of phase velocity and UA reconstructions.
For the numerical solution of the optimization problem Eq.\ref{eq:inv_forms}, a limited-memory Broyden-Fletcher-Goldfarb-Shanno (L-BFGS) algorithm~\cite{broyden1970convergence,fletcher1970new, goldfarb1970family, shanno1970conditioning} is used from the unconstrained optimization package \textit{minFunc}\footnote{https://www.cs.ubc.ca/~schmidtm/Software/minFunc.html}.

\subsection{Model fitting}
Ultrasound phase velocity $c(\omega)$ and attenuation $\alpha(\omega)$ reconstructions are computed
\begin{figure*}
    \begin{center}
    \includegraphics[width=0.85\textwidth]{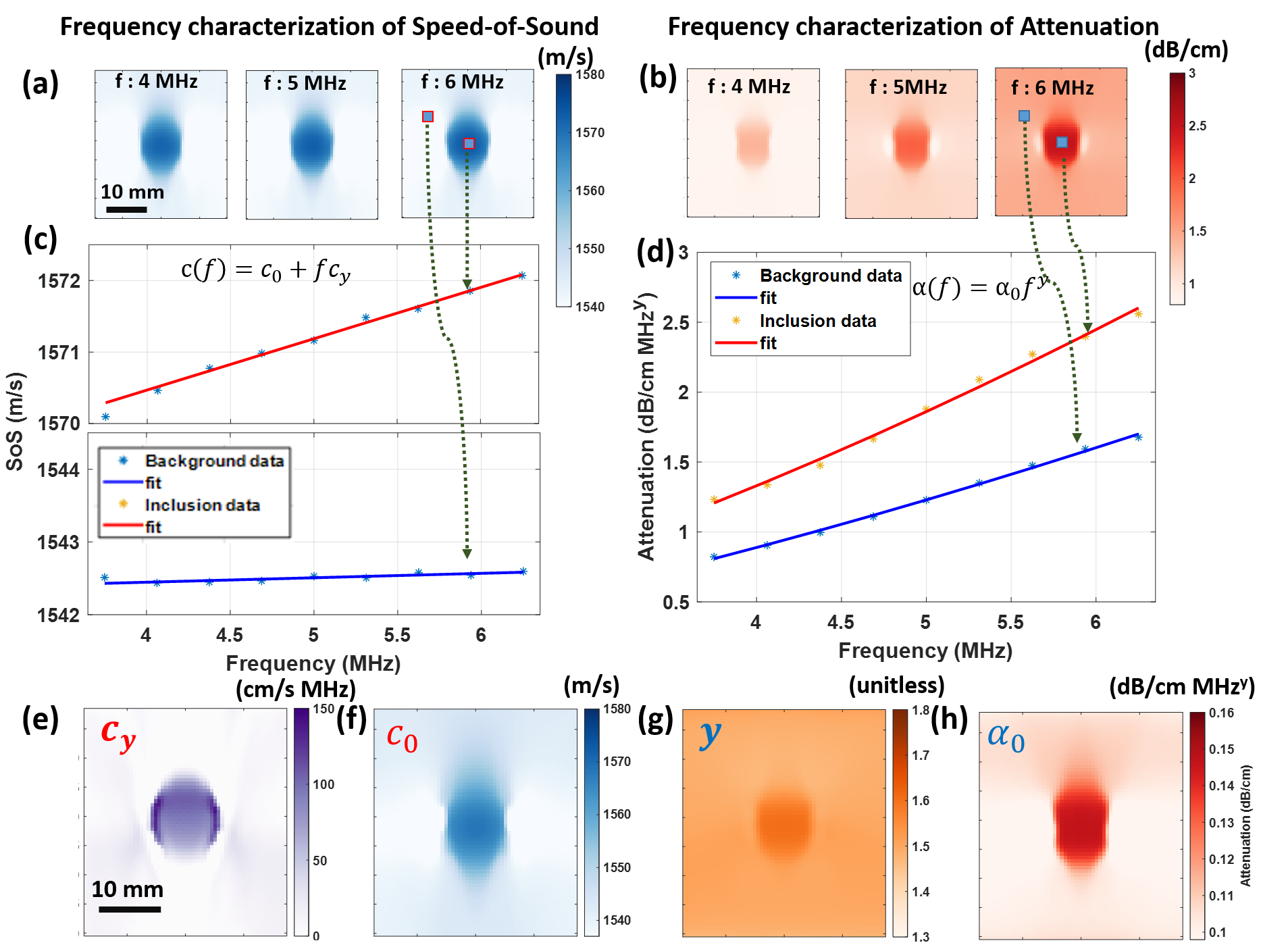}
    \caption{Illustration of parameter estimation for SoS and UA on a sample simulated numerical phantom: SoS~(a) and UA~(b) reconstructions at frequencies $f=\{4, 5, 6\}$\,MHz obtained by following the procedure in Fig.\,\ref{fig:SoS_UA_recon}.
    Fitting frequency-dependent models as in Eqs.\,\ref{eq:models} for SoS~(c) UA~(d) at two sample locations inside and outside the inclusion.
    By solving the inverse problem in Eq.\,\ref{eq:model_parameters_inv}, spatial distributions of the following model parameters are obtained: SoS coefficient $i.e.\ c_y$~(e), SoS dispersion coefficient $c_0$~(f), UA power-law exponent $y$~(g), and UA coefficient $\alpha _0$~(h). }
    \label{fig:model_fitting}
    \end{center}
\end{figure*}
for a range of frequencies $[\omega_c-n\Delta\omega,..., \omega_c, ..., \omega_c+n\Delta\omega]$ around the center frequency $\omega_c$ of the ultrasound pulse with $\Delta\omega$ being the frequency resolution used for computing the FT of the reflector profiles.
Sample reconstructions at three frequency values are shown in Fig.\,\ref{fig:model_fitting}(a-b).
Phase velocities for wide range of frequencies may exhibit complex profiles as in~\cite{waters2000applicability,waters2005causality,treeby2010modeling}.
Nevertheless, for small ranges of ultrasound imaging frequencies, such frequency dependence can be approximated with a linear disperse model, similarly to the shear-wave phase velocity studies in~\cite{barry2012shear,chintada2020nonlinear}.
Given the relatively small bandwidth of the utilized ultrasound pulse, the phase velocity is expected to vary linearly within this bandwidth, as illustrated by two sample profiles in Fig.\,\ref{fig:model_fitting}(c).
For frequency characterization of attenuation, it is well known in the literature that attenuation as a function of frequency obeys the \emph{power-law}, which corroborates the observation in  Fig.\,\ref{fig:model_fitting}(d).
One could fit these models pixel-wise for the given range frequencies and solve for the model parameters in the least square sense, as in~\cite{rau2021frequency}. 
However, this may be error prone as it ignores the spatio-spectral continuity in the medium. 
Herein, we incorporate frequency-dependent models of speed and attenuation into the reconstruction process, such that the inverse problem solution is informed by and can leverage the measurements at each frequency and path \emph{concurrently}, while estimating the frequency-dependent medium parameters.

As models of the frequency-dependent nature of ultrasound phase velocity and attenuation, we use herein the following linear and power-law relations,
respectively:
\begin{subequations}
    \label{eq:models}
    \begin{align}
        \label{eq:phase_velocity_model}
        c(f) &= c_0 + c_yf \\
        \label{eq:attenuation_model}
        \alpha(f)&= \alpha_0 f^y;
    \end{align}
\end{subequations}
where $c(f)$ and $\alpha(f)$ are the frequency-dependent SoS and attenuation, $c_0$ and $\alpha_0$ the medium-specific SoS and attenuation coefficients, $c_y$ the medium-specific SoS dispersion coefficient, and $y$ the medium-specific attenuation exponent.

By colocating SoS reconstructions for all measurement frequencies from  Eq.\,\ref{eq:inv_sos} in Eq.\,\ref{eq:phase_velocity_model}, a comprehensive system of equations that encodes both SoS model parameters for all image locations is arrived as follows:
\begin{equation}
     \label{eq:model_parameters_inv}
     \boldsymbol {\hat \theta} = \arg\min_{\boldsymbol\theta} {||A\boldsymbol\theta-\boldsymbol v||_1} + \lambda||D'\boldsymbol\theta||_1 \,
\end{equation}
where 
$\boldsymbol\theta =
\left[
{\bf c_0}\  
{\bf c_y}
\right]^\mathrm{T}
$\,, $\boldsymbol v= 
\left[
{\bf c_{f_1}}\ 
{\bf c_{f_2}}\ 
\cdots\ 
{\bf c_{f_n}}
\right]^\mathrm{T}$\,, 
$D'=[D\, D]$\,, and 
\[A =
\left[
\mkern-12mu
\begin{array}{cc}
I_{n_p} & f_1*I_{n_p}\\
I_{n_p} & f_2*I_{n_p}\\
\vdots & \vdots\\
I_{n_p} & f_{n}*I_{n_p}\\
\end{array}
\mkern-12mu
\right]
\]
with bold variables denoting size-$n_p$ row vectors of the sought parameters in the imaged field-of-view, $I_{n_p}$ is the identity matrix of size $n_p$ and $n_p$ is the total number of reconstruction grid pixels.

For estimating UA parameters, we colocate UA reconstructions for all measurement frequencies from  Eq.\,\ref{eq:inv_attenuation} in Eq.\,\ref{eq:attenuation_model}.
Taking the logarithm of both sides (in order to drop down the exponent) a similar linear problem as in Eq.\,\ref{eq:model_parameters_inv} is arrived, but this time with the variables 
$\boldsymbol\theta =
\left[
{\log\boldsymbol\alpha_0},\,
{\boldsymbol y}
\right]^\mathrm{T}
$\,, $\boldsymbol v=
\left[
\log\boldsymbol\alpha_{\bf{f_1}},
\log\boldsymbol\alpha_{\bf{f_2}},
\cdots,
\log\boldsymbol\alpha_{\bf{f_n}}
\right]^\mathrm{T}$\,, and
\[A=
\left[
\mkern-12mu
\begin{array}{cc}
I_{n_p} & \log(f_1)*I_{n_p}\\
I_{n_p} & \log(f_2)*I_{n_p}\\
\vdots & \vdots\\
I_{n_p} & \log(f_{n})*I_{n_p}\\
\end{array}
\mkern-12mu
\right]\ .
\]

Solving these inverse problem formulations we estimate the spatial maps of SoS model parameters $c_0$ and $c_y$, and of UA model parameters $\alpha_0$ and $y$.
Sample reconstructions from the earlier simulated example can be seen in Fig.\,\ref{fig:model_fitting}(e-h).

\subsection{Complex bulk modulus}
Given ultrasound phase velocity $c(\omega)$ and attenuation $\alpha(\omega)$ maps reconstructed above, complex bulk modulus $\hat K$ can be determined in a model-independent way to characterize the viscoelastic nature of the tissue through its relationship to complex wave number as follows:
\begin{equation}
    \label{eq:complex_ultrasound_wave_number}
     \hat{K}(\omega) = K'(\omega)+ iK''(\omega) =  \frac{\rho\omega^2}{\hat{k}^2(\omega)},
\end{equation}
where $\rho$ represents tissue density. 
Substituting Eq.\,\eqref{eq:ultrasound_propb} in Eq.\,\eqref{eq:complex_ultrasound_wave_number} allows to derive the bulk storage modulus $K'$ and bulk loss modulus $K''$ as follows:
\begin{subequations}
     \label{eq:complex_mod}
     \begin{align}
         \label{eq:bulk_storage_mod}
         K'(\omega) &= \rho \omega^2 \frac{\big(\frac{\omega}{c(\omega)}\big)^2-\alpha(\omega)^2}{\Big(\big(\frac{\omega}{c(\omega)}\big)^2+\alpha(\omega)^2\Big)^2}\,, \\
         \label{eq:bulk_loss_mod}
         K''(\omega) &= 2\rho \omega^2 \frac{\big(\frac{\omega}{c(\omega)}\big)\alpha(\omega)}{\Big(\big(\frac{\omega}{c(\omega)}\big)^2+\alpha(\omega)^2\Big)^2}\ . 
     \end{align}
\end{subequations}

\subsection{Evaluation metrics}
\label{sec:eval_metrics}
In this study, the following metrics are used for a quantitative analysis of the simulation results:

\begin{itemize}
    \item Root-mean-squared-error (RMSE)\\
        $ RMSE = \sqrt{ \frac{1}{n_p}{\left(x^\star-\hat x\right)}^2}$, where $x$ represents a reconstructed parameter map with the number of pixels $n_p$.
        The $\cdot^\star$ and $\hat{\cdot}$ indicate, respectively, the ground-truth and a reconstruction.
    \item Contrast-ratio fraction (CRF)\\
        $ CRF= \frac{\hat C}{C^\star}$, where $C=\frac{|\mu_\mathrm{inc} -\mu_\mathrm{bg}|}{|\mu_\mathrm{inc}|+|\mu_\mathrm{bg}|}$ with mean inclusion and background values $\mu_\mathrm{inc}$ and $\mu_\mathrm{bg}$, respectively, where the inclusion is delineated using the ground-truth map.
    \item Contrast-to-noise ratio (CNR)\\
        $CNR =\frac{|\mu_\mathrm{inc}|-|\mu_\mathrm{bg}|}{\sqrt{\sigma_\mathrm{inc}^2+\sigma_\mathrm{bg}^2}}$, where $\sigma^2$ represents the variance. 
\end{itemize}
 
\subsection{Simulation experiments}
\textit{Numerical simulations} were conducted to study the reconstruction accuracy of the proposed methods.
These simulations were conducted by varying local distribution of SoS and UA patterns and varying their frequency dependency characteristics.
An open-source acoustics toolbox k-Wave~\cite{treeby2010k} was used for this purpose.
k-Wave uses Kramers–Kronig relations~\cite{waters2000applicability,waters2005causality,treeby2010modeling} to simulate the dispersive phase speed based on given spatial maps of UA coefficient $\alpha_0(x,y)$ and SoS $c_{f_c}(x,y)$ at the center frequency, as well as the power-law dependency parameter $y$.
Note that in k-Wave, $y$ is a single constant for the entire domain/simulation and cannot be controlled spatially.

The transducer was simulated as a linear array of 128 elements with a pitch of 0.3\,mm.
Full-matrix multistatic data was simulated at a center frequency of 5\,MHz with a 5-cycle input pulse.
For the simulation, we used a temporal resolution of 160\,MHz and a spatial grid resolution of 37.5\,µm.
In the physical experimental setup, the acoustic mirror is made of plexiglass, so it was simulated using plexiglass's acoustic properties (density $\rho=1180$\,kg/m$^3$, SoS $c = 2700$\,m/s), as positioned at a distance of 42\,mm from the transducer surface in all three simulations.
A single  circular inclusion with 5\,mm radius is placed at the center of the imaging field-of-view.
SoS at center frequency ($c_{f_c}$) in the background and inclusion were set, respectively, 1540\,m/s and 1601.6\,m/s (for 4\% contrast).
The inclusion and background attenuation coefficient ($\alpha_0$) were set, respectively, 0.2\,dB/cm$\cdot$MHz$^\mathbf{y}$ and 0.1\,dB/cm$\cdot$MHz$^\mathbf{y}$.

The frequency characteristics of SoS and UA were varied across 
We repeated the simulations three times by varying the power exponent $y$= \{1.1, 1.5, 1.9\}, which led to three numerical phantoms referred hereafter as \emph{sim}\{\#1, \#2, \#3\}, respectively.
Note that given constraints in k-Wave, SoS and UA cannot be controlled independently and arbitrarily, and the variation of such exponent $y$ changes both SoS and UA with a known relation.

\subsection{Phantom and \textit{ex-vivo} experiments}
These were conducted using the data from~\cite{rau2021frequency}.
In the first phantom \#A we used, the background has 10\% gelatin 1\% Sigmacell Cellulose Type 50 (Sigma Aldrich, St. Louis, MO, USA), into which a bovine skeletal muscle sample was inserted as an inclusion.
For the second phantom \#B, the background and inclusion compositions were interchanged, i.e.\ a gelatin phantom piece was inserted as the inclusion inside a bulk muscle sample.
Data acquisitions were conducted on a research ultrasound machine (Verasonics, Kirkland, WA, USA) with a 128-element linear-array transducer (Philips, ATL L7-4). 
Similarly to the simulation setup, a wideband Tx pulse with a center frequency of 5.2\,MHz and a pulse length of 5 cycles was used.
For the calibration procedure, multi-static data sets were acquired in distilled water at room temperature by placing the reflector at multiple depths \{31, 35, 39, 43, 47\}\,mm.
For any intermediate reflector depths used in the phantom measurements, calibration data was interpolated from the available water measurements above, as described in~\cite{Rau_attenuation_19}.  

\section{Results and Discussion}

First, the reflector profiles are delineated in the multistatic data for each Tx event using~\cite{chintada2021time}.
To identify the acoustic reflector surface, we tuned this reflector delineation framework to use the so-called ``edge'' feature with a window length of 200 and 50, respectively, for simulations and \textit{ex-vivo} experiments. 
A cubic RANSAC~\cite{fischler1981random} model was used for outlier removal, to estimate an initial contour, which was next refined using an optimization-based active contours~\cite{kass1988snakes} method, which hence utilizes the expected temporal continuity of the neighbouring profiles for a robust estimation. 
This then yields the final delineation of reflector echoes to estimate the time-of-flights as well as the temporal echo profiles around the delineated reflector surface (to further process) as exemplified in Fig.\,\ref{fig:methods_1}c. 
This delineation procedure was repeated for all 128 Tx events.
For each delineated reflector profile of a Tx-Rx combination, 1-D temporal Fourier transform was performed using FFT at a resolution of 312.5\,kHz and 325.5\,kHz, respectively, for simulations and \textit{ex-vivo} experiments to extract phase and amplitude spectrums, as exemplified in Figs.\,\ref{fig:SoS_UA_recon}(a-b).
Repeating this for all Tx-Rx combinations yields 128$\times$128 spectrum profiles for phase and amplitude.
Next, phase and amplitude values at a frequency of interest are extracted and calibrated using the respective calibration parameters, derived separately for (simulated) numerical and physical experiments.
From the calibrated measurements for all Tx-Rx combinations, SoS and UA maps are reconstructed using Eqs.\,\ref{eq:inv_forms} and~\ref{eq:model_parameters_inv}.

\subsection{Simulation results}
\begin{figure}
    \centering
    \includegraphics[width=\columnwidth]{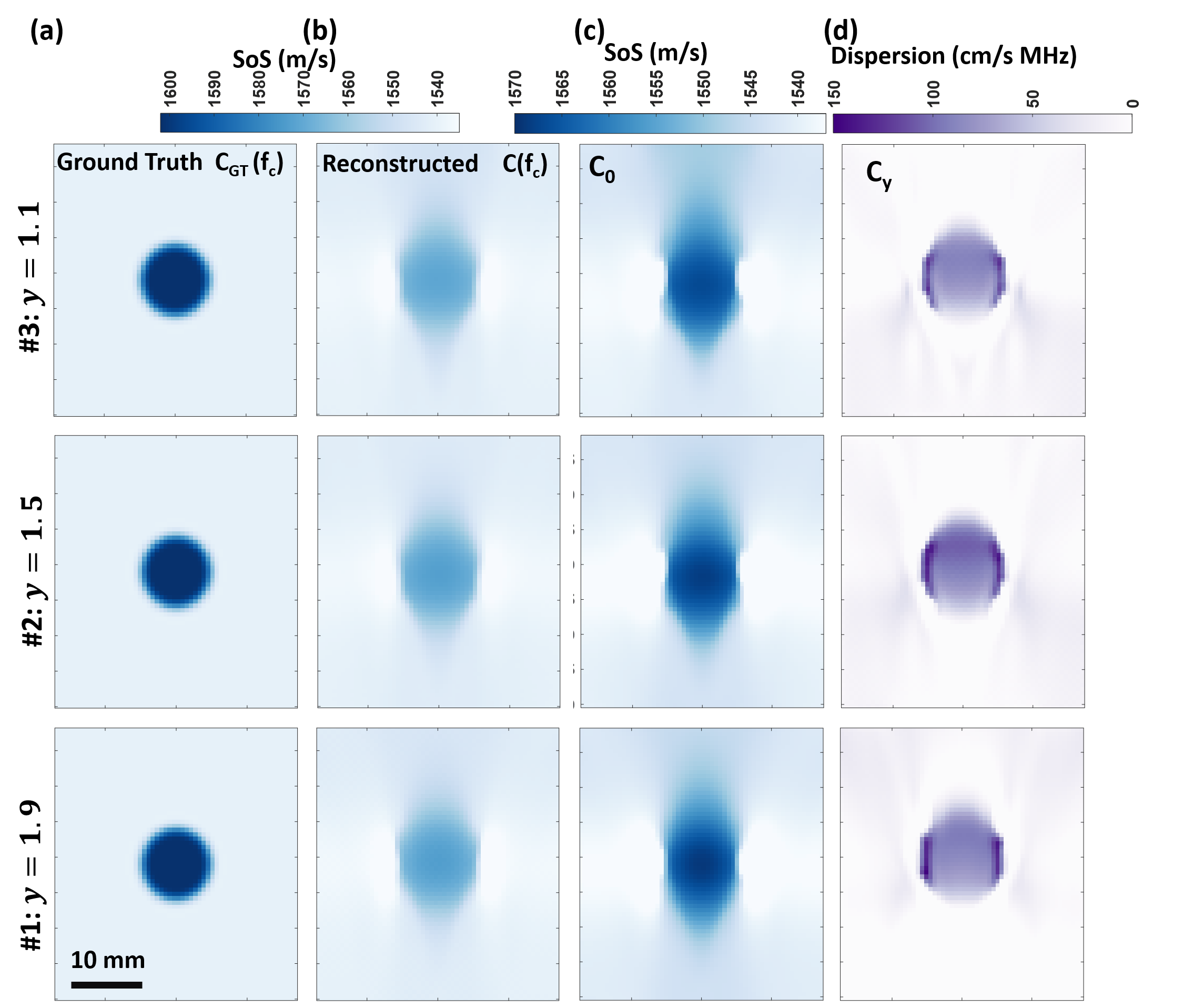}
    \caption{Comparison of frequency-dependent SoS imaging for the three different simulations with y=\{1.1, 1.5, 1.9\} in each row, with columns from left to right: ground-truth SoS maps $c_\mathrm{GT}({f_c})$ at the center frequency $f_c$=5\,MHz; reconstructed SoS maps $c({f_c})$ at the center frequency; reconstructed $c_0$ maps; and reconstructed $c_y$ maps.}
    \label{fig:simulations_sos_maps}
\end{figure}

SoS and UA reconstructions for \emph{sim}\#2 are presented in Fig.\,\ref{fig:SoS_UA_recon}.
Reconstructed $c_0$ (i.e., SoS at the center frequency $c({f_c})$) and the linear dispersion coefficient $c_y$ for the three simulations are depicted in Fig\,\ref{fig:simulations_sos_maps}. 
From the ground-truth and reconstructed SoS maps at the center frequency in Fig.\,\ref{fig:simulations_sos_maps}(a-b), it can be seen that the background SoS at the center frequency is reconstructed with some success in all cases.
However, the inclusion SoS values are seen to be shifted towards background values, which corroborates the findings in~\cite{sanabria2018speed}.
This is due to the limited-angle tomographic nature of the problem, where the lack of lateral projections combined with the regularization needed for robust solutions cause the smoothing and axial elongation of the inclusions, hence spreading them over a larger area. 
Since the cumulative SoS effect shall stay the same, per-pixel inclusion values are effectively averaged with the background, inversely proportional to such artifactual area increase.

\begin{figure}
    \centering
    \includegraphics[width=\columnwidth]{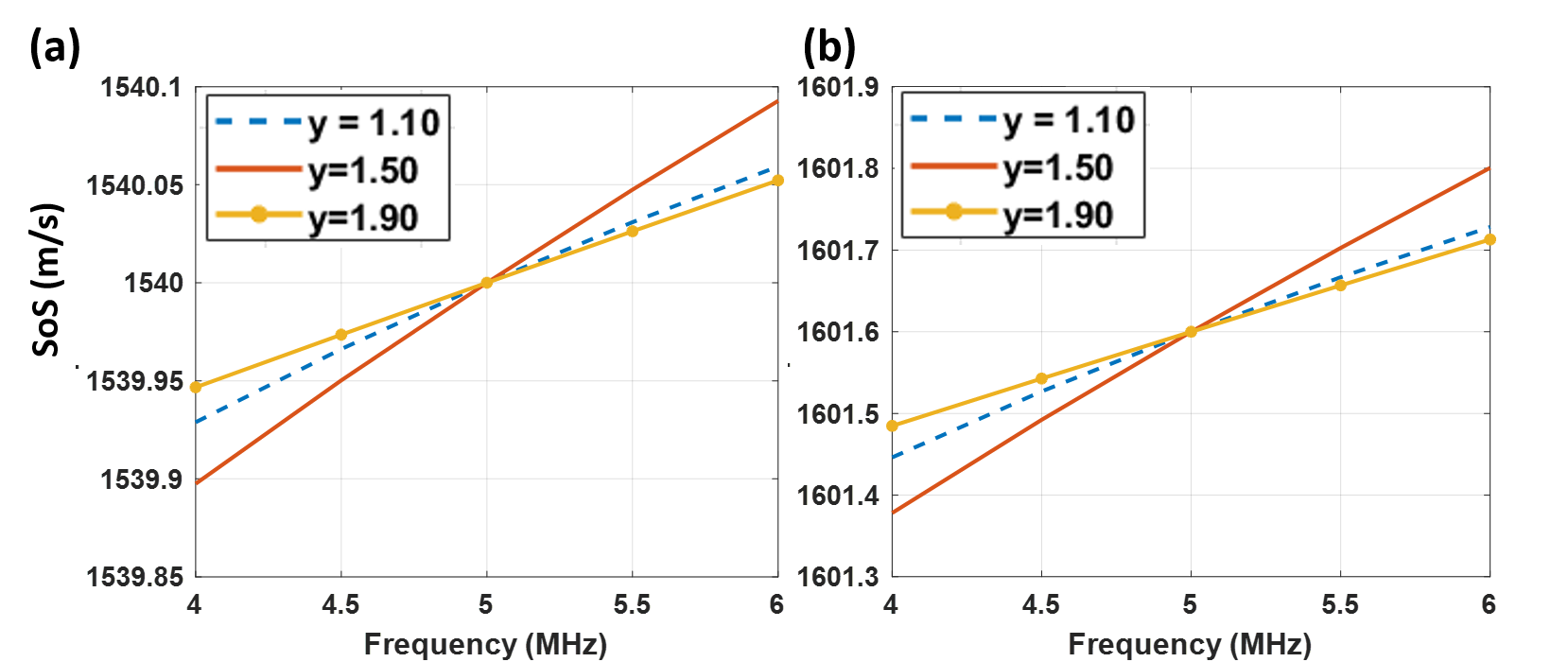}
    \caption{SoS dispersion characteristics for the background~(a) and inclusion~(b) for k-Wave simulations with power-law exponent $y=$\{1.1, 1.5, 1.9\} according to the Kramers-Kronig relationship that is being used in the simulation toolbox k-Wave~\cite{waters2000applicability,waters2005causality,treeby2010modeling}.}
    \label{fig:sos_dispersion_coefficients}
\end{figure}

Note that as $y$ varies across the simulations, so does $c_y$, according to Kramers–Kronig relationship~\cite{waters2005causality}. 
Therefore, the $c_y$ maps in Fig.\,\ref{fig:simulations_sos_maps}(d) are expected to present contrast between the three simulations with different $y$ values.
An analytical computation of expected SoS dispersion in the utilized frequency range of 4-6\,MHz according to the Kramers-Kronig relationship is illustrated in  Fig.\,\ref{fig:sos_dispersion_coefficients}.
Accordingly, in this frequency range a linear SoS dispersion ($c_y$) of \{6.5, 9.7, 5.3\}\,cm/s$\cdot$MHz for the background and \{14.1, 21.0, 11.4\}\,cm/s$\cdot$MHz for the inclusion are expected for each respective simulation. 
These are used as ground-truth $c_y$ values for the following evaluation.

For a quantitative evaluation, we computed the metrics RMSE, CNR, and CRF for each simulation, as reported in Tab.\,\ref{tab:simulation_results_tab}.
\renewcommand{\tabcolsep}{1.7pt}
\begin{table*}
    \caption{Quantitative evaluation of SoS and UA frequency-dependent model parameter reconstructions for simulations. $\mu_\mathrm{bg}$ and $\sigma_\mathrm{bg}$ are mean and standard deviation of background, $\mu_\mathrm{inc}$ and $\sigma_\mathrm{inc}$ are mean and standard deviation of inclusion.}
    \label{tab:simulation_results_tab}
    \begin{center}
    \small
    \begin{tabular}{|c||c|c|c|c|c||c|c|c|c|c||c|c|c|c|c||}
    \multicolumn{1}{c||}{\textbf{Sim}} & \multicolumn{5}{c||}{ $c(f_c)$\,(m/s) at $f_c=$\,5MHz} & \multicolumn{5}{c||}{$c_0$\,(m/s)} & \multicolumn{5}{c||}{Dispersion coefficient ($c_y$)\,(cm/s$\cdot$MHz)} \\  \multicolumn{1}{c||}{}
    & $\mu_{bg}$\scriptsize{$\pm\sigma_{bg}$} & $\mu_{inc}$\scriptsize{$\pm\sigma_{inc}$} & RMSE & CRF & CNR & $\mu_{bg}$\scriptsize{$\pm\sigma_{bg}$} & $\mu_{inc}$\scriptsize{$\pm\sigma_{inc}$} & RMSE & CRF & CNR & $\mu_{bg}$\scriptsize{$\pm\sigma_{bg}$} & $\mu_{inc}$\scriptsize{$\pm\sigma_{inc}$} & RMSE & CRF & CNR \\ \hline
    \multicolumn{1}{c||}{\#1} & 1542.4\scriptsize{$\pm$6.1} & 1568.2\scriptsize{$\pm$2.5} & 8.8 & 0.4 & 5.6 & 1542.0\scriptsize{$\pm$6.1} & 1564.4\scriptsize{$\pm$1.7} & 9.7 & 0.4 & 5.0 & 38.4\scriptsize{$\pm$22.8} & 74.5\scriptsize{$\pm$35.0} & 50.2 & 16.3 & 1.2 \\ 
    \multicolumn{1}{c||}{\#2} & 1542.2\scriptsize{$\pm$6.0} & 1569.8\scriptsize{$\pm$1.3} & 8.4 & 0.5 & 6.0 & 1541.7\scriptsize{$\pm$5.7} & 1565.4\scriptsize{$\pm$2.1} & 9.2 & 0.4 & 5.5 & 39.0\scriptsize{$\pm$24.6} & 87.3\scriptsize{$\pm$33.3} & 53.2 & 19.5 & 1.7 \\
    \multicolumn{1}{c||}{\#3} & 1542.2\scriptsize{$\pm$5.9} & 1569.5\scriptsize{$\pm$1.9} & 8.4 & 0.5 & 6.0 & 1542.0\scriptsize{$\pm$5.6} & 1565.5\scriptsize{$\pm$2.2} & 9.1 & 0.4 & 5.5 & 19.0\scriptsize{$\pm$24.8} & 79.3\scriptsize{$\pm$16.2} & 39.4 & 31.3 & 2.9 \\ \hline \hline
    \multicolumn{1}{c||}{\textbf{Sim}} & \multicolumn{5}{c||}{$\alpha(f_c)$\,(dB/cm) at $f_c=$\,5MHz} & \multicolumn{5}{c||}{UA coefficient ($\alpha _0$)\,(dB/cm$\cdot$MHz$^y$)} & \multicolumn{5}{c||}{UA exponent ($y$)\,(unitless)}  \\  \multicolumn{1}{c||}{}
    & $\mu_{bg}$\scriptsize {$\pm\sigma_{bg}$} & $\mu_{inc}$\scriptsize{$\pm\sigma_{inc}$} & RMSE & CRF & CNR & $\mu_{bg}$\scriptsize{$\pm\sigma_{bg}$} & $\mu_{inc}$\scriptsize{$\pm\sigma_{inc}$} & RMSE & CRF & CNR & $\mu$\scriptsize{$\pm\sigma$} & - & RMSE & - & - \\ \hline
    \multicolumn{1}{c||}{\#1} & 0.6\scriptsize{$\pm$0.1} & 1.0\scriptsize{$\pm$0.2} & 0.1 & 0.7 & 2.4 & 0.1\scriptsize{$\pm$0.01} & 0.17\scriptsize{$\pm$0.03} & 0.01 & 0.8 & 2.6 & 1.1\scriptsize{$\pm$0.03} & - & 0.04 & - & - \\ 
    \multicolumn{1}{c||}{\#2} & 1.1\scriptsize{$\pm$0.1} & 1.8\scriptsize{$\pm$0.2} & 0.1 & 0.7 & 4.9 & 0.1\scriptsize{$\pm$0.01} & 0.14\scriptsize{$\pm$0.01} & 0.01 & 0.5 & 4.5 & 1.5\scriptsize{$\pm$0.05} & - & 0.05 & - & -\\
    \multicolumn{1}{c||}{\#3} & 2.2\scriptsize{$\pm$0.2} & 3.4\scriptsize{$\pm$0.3} & 0.3 & 0.6 & 4.3 & 0.1\scriptsize{$\pm$0.01} & 0.15\scriptsize{$\pm$0.01} & 0.01 & 0.5 & 4.2 & 1.9\scriptsize{$\pm$0.06} & - & 0.06 & - & - \\ \hline 
    \end{tabular}
    \normalsize
    \end{center}
\end{table*}
We computed the mean and standard deviation for the background and inclusion using their respective masks from simulations.
The average RMSE of $c(f_c)$ is 8.50 $\pm$ 0.24\,m/s and the average mean values of $c({f_c})$ in the background region is 1542.3$\pm$0.1\,m/s, demonstrating that an accurate reconstruction of the speed of sound is possible.
The contrast metrics CNR and CRF for $c(f_c)$ and for $c_0$ indicate that the inclusions can be successfully distinguished in SoS reconstructions, with a contrast similar to their prescribed ground-truth value.
Note that $c_y$ reconstructions for all the simulations show large RMSE errors of 39 to 53\,cm/s$\cdot$MHz.
This is mainly due to the relatively minute dispersion values of [5.3 - 9.7]\,cm/s$\cdot$MHz and hence their minimal effect in SoS change within the given frequency interval (see the changes in y-axis in Fig.\,\ref{fig:sos_dispersion_coefficients} being $<$0.4\,m/s for inclusion $<$0.2\,m/s for the background).
Note that this is partly due to our earlier-mentioned experimental limitations with k-Wave parameterizations, where SoS frequency dispersion can only only be generated by varying $y$, which then has a much more prominent impact on UA.
To capture such variation, very high SNR and precision would be required in SoS reconstructions. 
Despite the accuracy being low for these, nevertheless, the relative differences and contrasts are recovered successfully:
For instance, $c_y$ for the inclusion of sim \#2 is higher than that of sim \#1 and \#3, which corroborates the analytical expectation from Fig.\,\ref{fig:sos_dispersion_coefficients}. 
Similarly, corroborating the analytical expectation, the inclusions in sim \#1 and \#2 show roughly twice the dispersion that their backgrounds. 
The average ratio of $\mu^{c_y}_{inc}$ to $\mu^{c_y}_{bg}$ is $\approx$2.4, which is close to the ground truth $c_y$ contrast of 2 between inclusion and background.
Therefore, despite not being accurate 
reconstructed $c_y$ values are still correlated with ground truth counter-parts and hence may potentially act as an imaging biomarker. 

\begin{figure}
    \centering
    \includegraphics[width=\columnwidth]{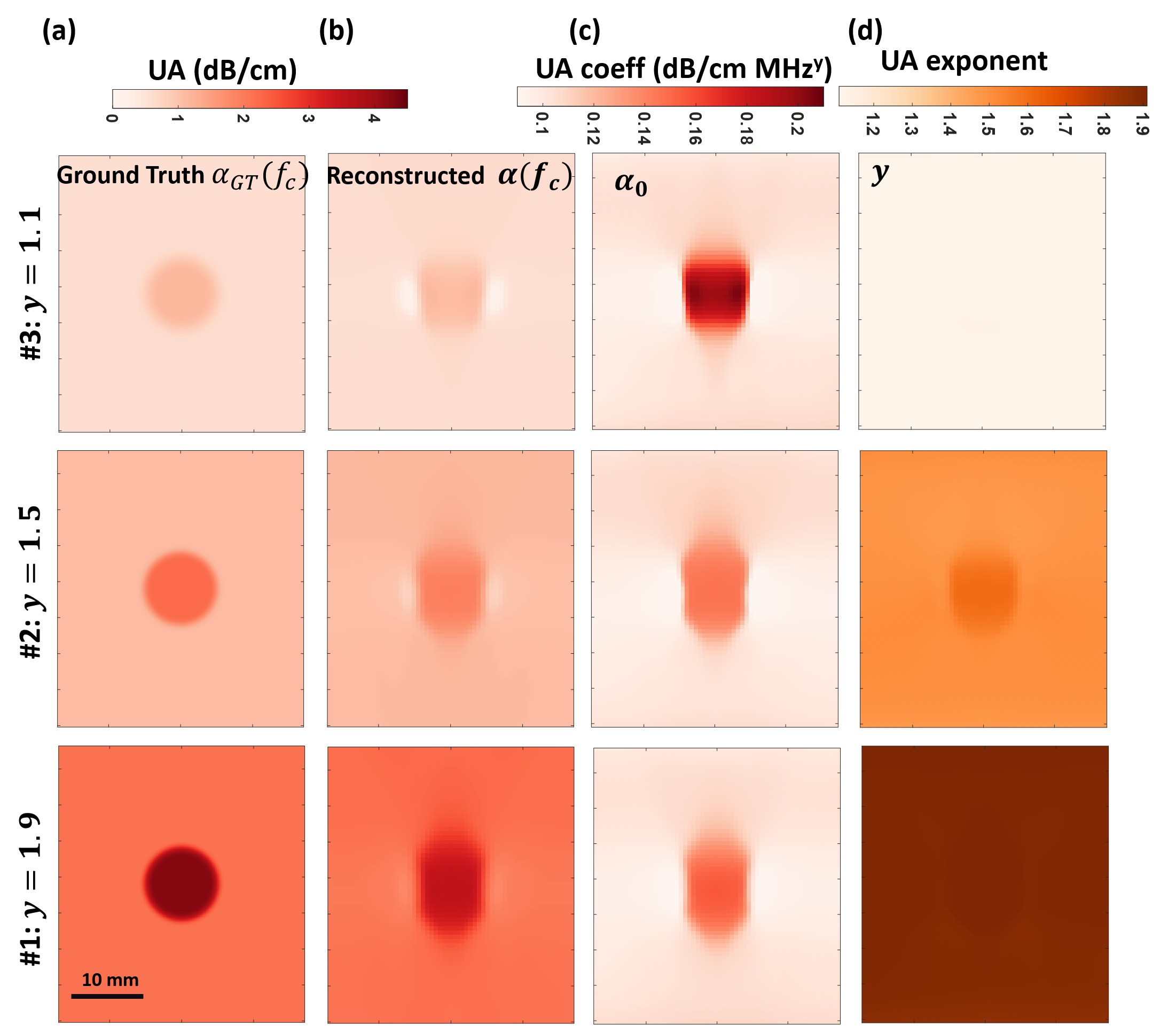}
    \caption{Comparison of frequency-dependent UA imaging for simulations varying the power-law dependence parameter y=\{1.1, 1.5, 1.9\}, with columns from left to right: ground-truth UA maps at center frequency 5\,MHz; reconstructed UA maps at the center frequency; reconstructed $\alpha_0$ maps;
    reconstructed $y$ maps.}
    \label{fig:simulations_alpha_maps}
\end{figure}

Reconstructed UA at the center frequency $\alpha_{fc}$, UA coefficient $\alpha_0$, and exponent $y$ are plotted in Fig.\,\ref{fig:simulations_alpha_maps}. 
Given the groundtruth and reconstructed $\alpha_{fc}$ maps in Fig\,\ref{fig:simulations_alpha_maps}(a-b), it can be seen that the UA at center frequency is successfully reconstructed in all cases.
In Fig.\,\ref{fig:simulations_alpha_maps}(b-c) it is observed that the reconstructed $\alpha_{fc}$ and  $\alpha_0$ values in the inclusions are shifted towards the background values, similarly to SoS results in Fig.\,\ref{fig:simulations_sos_maps}(b) and also in line with previous observations in~\cite{rau2021frequency}.
The UA exponent $y$ maps in Fig.\,\ref{fig:simulations_alpha_maps}(c) should not present any contrast since the power exponent $y$ are constant across each domain.
Although this is true for sim \#1, the other two simulations exhibit a slight deviation of $y$ in the inclusion region.
This is due to the relatively more pronounced artifactual elongations of the inclusion in UA reconstructions with increased frequency dispersion, which then slightly reduce the per-pixel reconstructed UA value (through averaging as reasoned above), which in turn is erroneously attributed by the model fitting to an increased $y$ in the inclusion.

To quantify UA reconstructions, we report RMSE, CNR, and CRF in Tab.\,\ref{tab:simulation_results_tab}.
Average RMSE of $\alpha(f_c)$ is 0.16$\pm$0.09\,dB/cm at center frequency 5\,MHz.
Considering the range of ground-truth SoS values, this demonstrates that an accurate reconstruction of UA is possible.
With an RMSE of \{0.04, 0.05, 0.06\} respectively for prescribed $y$ values \{1.1, 1.5, 1.9\}, the relative estimation error becomes below 3.3\%, indicating a robust  estimation of frequency exponent $y$.
The contrast metrics CNR and CRF for $\alpha({f_c})$, $\alpha _0$, and $y$ demonstrate that the inclusions can be successfully distinguished in UA reconstructions, with a contrast similar to their prescribed ground-truth values.

\subsection{\textit{Ex-vivo} experiments results}
\begin{figure}
    \centering
    \includegraphics[width=\columnwidth]{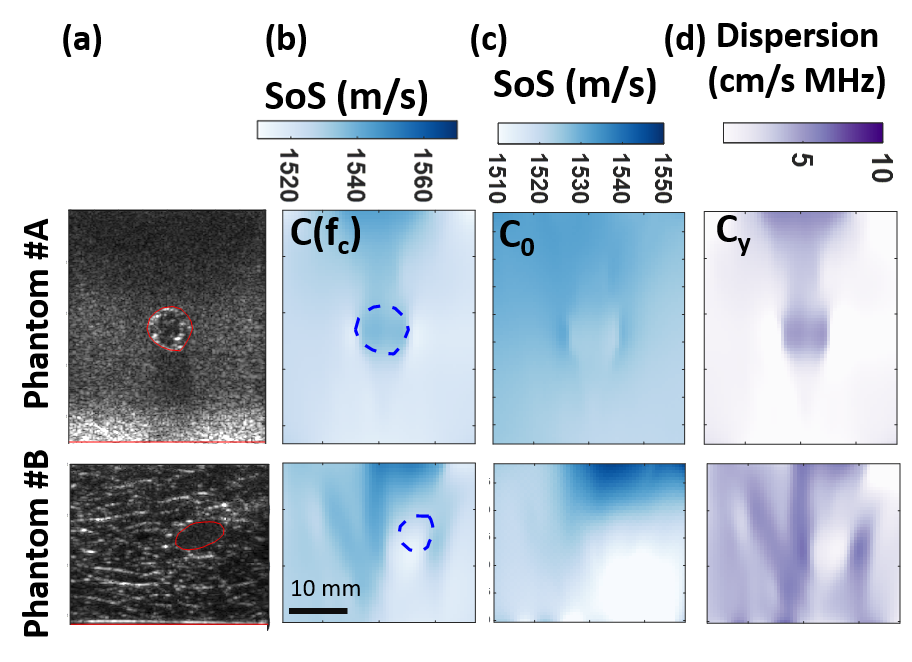}
    \caption{Comparison of frequency-dependent SoS imaging for phantom and \textit{ex-vivo} tissues, with columns from left to right: B-mode image; $c$ at the center frequency 5.2\,MHz; reconstructed $c_0$ maps; reconstructed $c_y$ maps.}
    \label{fig:ex_vivo_exps_sos_maps}
\end{figure}

The SoS reconstruction results of the gelatin phantom and \textit{ex-vivo} experiments are shown in Fig.\,\ref{fig:ex_vivo_exps_sos_maps}
with their corresponding B-Mode images in Fig.\,\ref{fig:ex_vivo_exps_sos_maps}(a).
Reconstructions of SoS maps at the center frequency are shown in Fig.\,\ref{fig:ex_vivo_exps_sos_maps}(b), while reconstructed $c_0$ and $c_y$ are shown in Fig.\,\ref{fig:ex_vivo_exps_sos_maps}(c\&d).
The reconstructed results in Fig.\,\ref{fig:ex_vivo_exps_sos_maps}(b) were overlaid with inclusion markings annotated from the B-mode images.
A clear contrast is visible in Fig.\,\ref{fig:ex_vivo_exps_sos_maps} for all our proposed reconstructions.
The reconstructed maps $c({f_c})$ and $c_y$ for gelatin and bovine tissue regions are consistent in phantoms \#A and \#B; i.e. the reconstructions show inverted images between these two phantoms, indicating that these quantities can be reproducibly used for tissue differentiation.
From Fig.\,\ref{fig:ex_vivo_exps_sos_maps}(c), it is observed that the tissue samples show much higher dispersion than in the simulations.
This indicates the selection of frequency for SoS quantification being an important factor in practice.
This also signifies the need for imaging $C_0$ and $C_y$ to image the SoS of the targeted medium comprehensively.
In other words, UA dispersion estimation can help both as additional parameters to characterize tissues and also for disambiguation of SoS measurement, which are seen here to be highly frequency dependent; also corroborating tissue observations in the literature, e.g.~\cite{bamber1981acoustic,duck2013physical}.  

\begin{figure}
    \centering
    \includegraphics[width=\columnwidth]{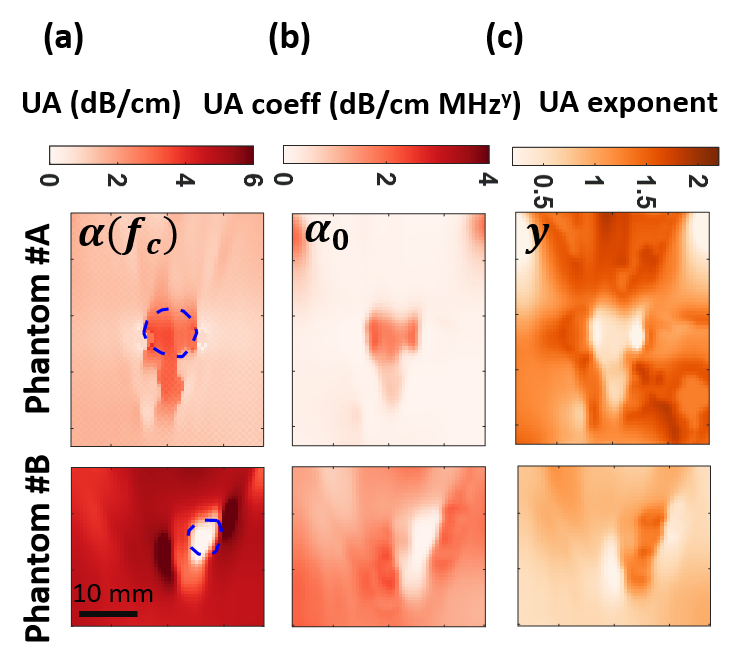}
    \caption{Comparison of frequency-dependent UA imaging for phantom and \textit{ex-vivo} tissues, with columns from left to right: $\alpha$ at the center frequency 5.2\,MHz; reconstructed $\alpha_0$ maps; reconstructed $y$ maps.}
    \label{fig:ex_vivo_exps_ua_maps}
\end{figure}

The UA results of the gelatin phantom and \textit{ex-vivo} experiments are shown in Fig.\,\ref{fig:ex_vivo_exps_ua_maps}.
A clear contrast is visible for all the UA reconstructions using our proposed methods.
Furthermore, all three reconstructed parameters between phantoms \#A and \#B are consistent, i.e.\ inclusion and background values inverted, showing reproducibility under different experimental settings.
Axial edge artifacts, especially in $y$ and $\alpha_0$ maps were much reduced compared to results in~\cite{rau2021frequency} thanks to solving model parameters collectively in a closed form as in Eq.\,\ref{eq:model_parameters_inv}.
The average and standard deviation values of SoS and UA obtained in gelatin and bovine tissue were tabulated along with CNR of the reconstructions in Tab.\,\ref{tab:ph_ex_vivo_results}.
\begin{table*}
    \caption{Quantitative evaluation of SoS and UA model parameter reconstructions for \textit{ex-vivo} experiments.}
    \label{tab:ph_ex_vivo_results}
    \begin{center}
    \begin{tabular}{|c||c|c|c||c|c|c||c|c|c||}
      \multicolumn{1}{c||}{\textbf{Phantom}} & \multicolumn{3}{c||}{{$c({f_c})$}\,(m/s) at $f_c=$\,5.2\,MHz} & \multicolumn{3}{c||}{$c_0$\,(m/s)} & \multicolumn{3}{c||}{$c_y$\,(cm/s$\cdot$MHz)}\\ \multicolumn{1}{c||}{}
    & Gelatin & Tissue & CNR & Gelatin & Tissue & CNR & Gelatin & Tissue & CNR  \\ \hline
    \multicolumn{1}{c||}{\#A} & 1519.8\scriptsize{$\pm$06.69} & 1535.5{\scriptsize$\pm$03.11} & 3.00 & 1526.5\scriptsize{$\pm$3.60} & 1525.1\scriptsize{$\pm$01.50} & 0.51 & 100.0\scriptsize{$\pm$165.0} & 430.0\scriptsize{$\pm$101.0} & 2.41\\ 
    \multicolumn{1}{c||}{\#B} & 1522.7\scriptsize{$\pm$06.82} & 1535.7\scriptsize{$\pm$12.85} & 1.27 & 1518.1\scriptsize{$\pm$7.22} & 1520.1\scriptsize{$\pm$11.59} & 0.21 & 152.0\scriptsize{$\pm$682.0} & 372.0\scriptsize{$\pm$698.0} & 0.32\\ \hline \hline 
    \multicolumn{1}{c||}{\textbf{Phantom}} & \multicolumn{3}{c||}{$\alpha({f_c})$ (dB/cm) at $f_c=$5.2MHz} & \multicolumn{3}{c||}{UA coefficient $\alpha _0$\,(dB/cm$\cdot$MHz$^y$)} & \multicolumn{3}{c||}{UA exponent $y$ (unitless)}  \\  \multicolumn{1}{c||}{}
    & Gelatin & Tissue & CNR & Gelatin & Tissue & CNR & Gelatin & Tissue & CNR\\ \hline
    
    \multicolumn{1}{c||}{\#A} & 1.39\scriptsize{$\pm$0.38} & 2.95\scriptsize{$\pm$0.40} & 4.02 & 0.24\scriptsize{$\pm$0.32} & 1.43\scriptsize{$\pm$0.61} & 2.44 & 1.26\scriptsize{$\pm$0.39} & 0.54\scriptsize{$\pm$0.29} & 2.09\\
    \multicolumn{1}{c||}{\#B} & 1.73\scriptsize{$\pm$1.58} & 4.69\scriptsize{$\pm$0.89} & 2.30 & 0.19\scriptsize{$\pm$0.40} & 1.43\scriptsize{$\pm$0.75} & 2.06 & 1.30\scriptsize{$\pm$0.25} & 0.76\scriptsize{$\pm$0.31} & 1.91\\ \hline
    \end{tabular}
    \end{center}
\end{table*}
Using the relationship in Eq.\,\ref{eq:complex_mod}, the complex bulk modulus of the gelatin phantom and \textit{ex-vivo} tissues at ultrasound center frequency can be computed to be $2.31+i(2.42$$\times$$10^{-3})$\,GPa and $2.33+i(10.01$$\times$$10^{-3})$\,GPa, respectively.
These values indicate (1)~that the bulk loss modulus in both media is 3 orders of magnitude smaller than the bulk storage modulus; and (2)~that although both media have similar bulk storage modulus, the bulk loss modulus in \textit{ex-vivo} bovine tissue is $\approx$4 times higher than that of gelatin.
This striking difference indicates that the complex bulk modulus may act as a new imaging biomarker, where the loss modulus component potentially having superior tissue differentiation compared to its storage component.

\section{Conclusions}
In this study, a novel method for reconstructing local SoS and UA maps as a function of frequency is introduced, through frequency domain analysis followed by a closed-form fitting of linear SoS and power-law UA dependency models.
We first delineate the reflector profiles in multi-static data and compute Fourier transforms of these profiles to estimate phase and amplitude spectra. 
After calibrating these with water measurements, we use them to solve closed-form inverse problems over all frequencies to reconstruct SoS and UA frequency-dependent model parameters.
We have studied these reconstructions with simulations and \textit{ex-vivo} phantom studies, with our results indicating the following observations:
The introduced SoS model parameters $c_(f_c)$ (SoS at center frequency), $c_0$ (SoS intercept), and  $c_y$ (SoS dispersion) were reconstructed with average RMSEs of 8.50$\pm$0.24\,m/s, 9.30$\pm$0.31\,m/s, and 47.6$\pm$7.3\,cm/s$\cdot$MHz, respectively.
These results show that $c(f_c)$ and $c_0$ can be reconstructed with high accuracy, while $c_y$ reconstructions were relatively poorer within the simulated low-dispersion regime.
Still, the reconstructed $c_y$ maps and their contrasts correlated with the ground-truth values. 
UA model parameters $\alpha(f_c)$, $\alpha_0$, and $y$ were reconstructed with average RMSEs of 0.16$\pm$0.1\,dB/cm, 0.01$\pm$0dB/cm$\cdot$MHz$^\mathbf{y}$, and 0.05$\pm$0.01, respectively, which indicate a high overall reconstruction accuracy using our proposed methods. 
From phantom and \textit{ex-vivo} tissue sample experiments, the attenuation exponents in a gelatin-cellulose mixture and an ex-vivo bovine muscle sample were found to be, respectively, 1.3 and 0.6 on average.
Linear dispersion of SoS in a gelatin-cellulose mixture and an \textit{ex-vivo} bovine muscle sample were found to be, respectively, 1.3 and 4.0\,m/s$\cdot$MHz on average.
Bulk loss modulus in bovine muscle sample was $\approx$4 times the bulk loss modulus in the gelatin-cellulose mixture.
Our results show the feasibility of estimating spatial maps of frequency-dependent characteristics of SoS and UA, as well as the complex bulk modulus as potential imaging biomarkers.
\section{Acknowledgements}
Funding was provided by the Swiss National Science Foundation (SNSF).
\bibliographystyle{IEEEtran}
\bibliography{Bibilography.bib}
\end{document}